\documentclass[letterpaper,twocolumn,10pt]{article}
\usepackage{usenix2019_v3}

\usepackage{tikz}
\usepackage{amsmath}
\usepackage{xspace}
\usepackage{pgfplots}
\pgfplotsset{compat=1.9}

\usepackage{ntheorem}
\theorembodyfont{\upshape}

\usepackage{url}

\usepackage{filecontents}
\newcommand{\hash}{\texttt{TX-HASH}\xspace}
\newcommand{\id}{\texttt{TXID}\xspace}
\newcommand{\ids}{\texttt{TXIDs}\xspace}
\newcommand{\hashes}{\texttt{TX-HASHes}\xspace}
\newcommand{\name}{Txilm\xspace}

\newcommand*{\affaddr}[1]{#1} 
\newcommand*{\affmark}[1][*]{\textsuperscript{#1}}

\newtheorem{definition}{Definition}

\begin{document}

\date{}

\title{\Large \bf Txilm: Lossy Block Compression with Salted Short Hashing}



\author{%
\upshape{Donghui Ding}\affmark[1,2], 
\upshape{Xin Jiang}\affmark[1,2], 
\upshape{Jiaping Wang}\affmark[3,1], 
\upshape{Hao Wang}\affmark[3], 
\upshape{Xiaobing Zhang}\affmark[3],
\upshape{Yi Sun}\affmark[1]\\
\affaddr\upshape{{\affmark[1]Institute of Computing Technology, Chinese Academy of Sciences}}\\
\affaddr\upshape{{\affmark[2]University of Chinese Academy of Sciences}}\\
\affaddr\upshape{{\affmark[3]Monoxide Dev Team}}%
}

\maketitle

\begin{abstract}
 Current blockchains are restricted by the low throughput. Aimed at this problem, we propose Txilm, a protocol that compresses the size of transaction presentation in each block to save the bandwidth of the network. In this protocol, a block carries short hashes of \ids instead of complete transactions. Combined with the sorted transactions based on \ids, Txilm realizes 80 times of data size reduction compared with the original blockchains. We also evaluate the probability of hash collisions, and provide methods of resolving such collisions. Finally, we design strategies to protect against potential attacks on Txilm. 
\end{abstract}

\section{Introduction}
\label{Sec:Intro}

Blockchains have been applied to wide areas such as cryptocurrencies and finance. However, current blockchains are restricted by the low throughput. For example, Bitcoin only handles 7 transactions each second, while PayPal achieves 500 transactions/sec throughput and Visa even 4000 transactions/sec\cite{Kokoris2016Enhancing}.

Aimed at the throughput problem of blockchains, various schemes are proposed including the compact blocks\cite{BIP152, Xthin} in Bitcoin. A compact block carries only 32-byte \ids instead of complete transactions (300$-$400 bytes roughly), and the network bandwidth consumed by each transaction is thus reduced by around 10 times. Compact blocks are safe because most transactions have already been stored in the memory pool of each Bitcoin node.

On the basis of compact blocks, we propose Txilm, which uses the short hash of \ids to represent the transactions in the blockchain. This further reduces the size of the transaction representation in each block to around 40 bits. Txilm is simple without using additional data structures such as bloom filters or IBLTs (Invertible Bloom Lookup Tables)\cite{IBLT}. Furthermore, our proposal doesn't rely on consistent memory pools across full nodes.

We also optimize Txilm by sorting transactions based on \ids. Such optimization reduces the transaction representation in a block to 32 bits, which yields 8$\times$ compression over the original proposal of compact blocks. An 80 times of data size reduction is thus realized.

\section{Background}
\label{Sec:Background}
We now provide necessary background of this work including the details of the blockchain systems, and the throughput of blockchains.

\subsection{Blockchain System}
\label{SubSec:System}

A blockchain is a distributed ledger that records all updates of the system state. Each operation of the update is known as a $transaction$. Multiple transactions are packed into a block, and blocks are linked with each other to construct a complete blockchain.

A blockchain system is maintained by many full nodes. Each full node holds a consistent replica of the blockchain. miner is a special full node, which creates new blocks to extend the blockchain. Different nodes are connected with each other and form the blockchain network.

Bitcoin is a typical blockchain. A transaction in Bitcoin transfers assets from one user to another. A transaction is broadcast into the blockchain network, and propagated among Bitcoin nodes using the gossip protocol. When a miner receives a new transaction, the miner will put it into a memory pool, which is a list used to temporarily store newly-received transactions. When creating a new block, the miner will select some transactions from the memory pool, and packed them into the block. The size of a block is 1$-$1.2 MB, and usually carries 1000$-$3000 transactions \cite{size}.

Different miners create a new block concurrently. Bitcoin introduces a mathematical puzzle (i.e., proof of work, PoW) to decide which block is appended to the blockchain. Miners compete with each other, and the first miner who solves the puzzle will broadcast its block into the network. When a full node receives this block, it will verify the solution and then append this block to the blockchain.

Due to the propagation delay, other miners may also work out the solutions before this new block are broadcast into the entire network. This will cause inconsistent ledgers among different full nodes because each full node may receive new blocks in different orders. This temporary inconsistency is known as a $fork$. Bitcoin uses Longest-Chain Rule to determine the canonical chain. Since a fork can compromise the security of Bitcoin, the difficulty of PoW is dynamically adjusted to control the frequency of new block generation. Currently a new block is generated every 10 minutes.

Our work is based on Bitcoin, but can also be applied to Ethereum and other blockchains.

\subsection{Blockchain Throughput}
\label{SubSec:throughput}

The blockchain throughput, also known as TPS (Transactions Per Seconds), is the number of transactions recorded into the blockchain per unit time. Previous work \cite{TPS} has revealed an estimating formula of the throughput.

 \begin{equation}\label{equ:TPS}
  TPS = \beta\left(\lambda \right) \cdot b
\end{equation}

In Equation~\ref{equ:TPS}, $b$ is the number of transactions carried by a block, and $\beta$ is the rate of block addition to the canonical chain, which is positively related to the frequency of new block generation $\lambda$. Therefore, the throughput can be improved by:

\begin{enumerate}
\item Increasing $b$. We can compress the transaction size in each block, so a block can carry more transactions without causing pressure on the network bandwidth.

\item Improving $\lambda$. This can be achieved by reducing the difficulty of PoW. In order to reduce the probability of forks, the propagation delay of new blocks must be shortened. Research has shown that small-sized blocks can save the network bandwidth and thus shorten such delay \cite{propagation, scaling}. We can also compress the transaction size to construct smaller blocks.
\end{enumerate}

In conclusion, the compression of transaction size will save the network bandwidth, and is an effective method of improving the blockchain throughput. 
\section{Related Work}
\label{Sec:relatedWork}
Xthin \cite{Xthin} is the first scheme of transaction compression. A block in Xthin carries 256-bit \ids instead of complete transactions. The size of each block is thus reduced by around 10 times. BIP152 \cite{BIP152} proposes compact blocks to compress the block size in Bitcoin, which uses the same method as Xthin. Such schemes dramatically accelerate the propagation of new blocks in the network and therefore improve the blockchain throughput. However, subsequent research has shown that there is still a large compression space for the block size.

On the basis of Xthin, Xtreme \cite{Xtreme} and Graphene \cite{Graphene} further compress a Bitcoin block into 10 KB and 2.6 KB respectively. However, these schemes utilize additional data structures including bloom filters or IBLTs. This adds the complexity to the implementations. Xthinner \cite{Xthinner} uses the prefixes of \ids to represent transactions in each block, but it relies on the stack-based state machine to distinguish prefixes with unequal lengths.

\section{Txilm Protocol}
\label{Sec:Txilm}
In this section, we first introduce the design of \name Protocol, and then model the probability of hash collisions. We further reduce this probability using the sorted transactions based on \ids. And finally, we simulate hash collisions to evaluate our probability models.

\subsection{Protocol Design}
\label{SubSec:design}

Txilm derives from the compression of \ids. The 32-byte \id is the SHA256 value of a transaction, which acts as a unique identifier of this transaction in the Bitcoin blockchain. Based on the \id, the representation of a transaction can be further compressed by a short hash function:
\begin{center}
 \texttt{TXID-HASH} = \texttt{hash(TXID)}
\end{center}

In the above equation, \texttt{hash} is a function that generates 32-bit to 64-bit hash values, such as CRC32, CRC40 or CRC64. Each new compact block includes only a list of \hashes, ordered by the original list of transactions.

Ambiguity may occur with such a $k$-bit small-sized hash, which needs to be resolved by each full node. Once receiving a new block that includes the \hash list from the sender, the receiver searches each received \hash in the \id list produced by its memory pool. For each \hash, three cases may happen as follows:
\begin{enumerate}
\item Not found: There is no transaction in the memory pool that matches the received \hash. The receiver will ask the sender or other peers for the missing \id.
\item A single match is found: the \id for the received \hash is resolved.
\item Multiple matches are found: the receiver will collect all matched \ids as candidates for a 2nd-stage resolving.
\end{enumerate}

Each block header carries the SHA256-Merkle root of all contained \ids. In the 2nd-stage, all combinations of candidates are iterated to recompute this Merkle tree. A correct combination will result in a matched \id Merkle root with the one carried by the block header.

An optimization for the 2nd-stage is to add a lightweight pre-check before actually recomputing the Merkle root. We propose a lightweight Merkle tree for pre-check by replacing SHA256 with CRC32, which leads to a 4-byte root. When creating a new block, the 4-byte CRC32-Merkle root is computed using the \ids and encoded into the block. This yields a 40$\times$ acceleration in searching the right combination.

If any of the combinations can not match the Merkle root in the block header, the receiver will fall back to ask the sender to transfer the complete \id list of the block, which is described in the original BIP152 proposal. This situation happens when at least one \id in the memory pool has the same \hash in the received \hash list, but the \id is not the one that the block intends to include.

\subsection{Probability of a Single Collision}
\label{SubSec:Collision}

Iterating the combination of candidates consumes a considerable amount of CPU time, and incurs additional latency. The feasibility of \name thus highly relies on the probability of hash collisions. We use the \textbf{single collision} to model this probability.

\begin{definition}[Single Collision]
\label{def:sc}
We assume $M$ is the set of all unconfirmed transactions contained in a memory pool, and $b$ is an unconfirmed block. Given \hash $h \in b$ and its original transaction $tx_1$, $\exists$ $tx_2 \in M \vee tx_2 \in b$, $tx_1 \neq tx_2$, but \texttt{hash($tx_2$)}= $h$.
\end{definition}

In definition~\ref{def:sc}, a single collision can occur both within $b$ or between $b$ and $M$. $P_{sc}$ can be formulated as the Generalized Birthday Problem\cite{Birthday}. We assume a memory pool $M$ contains $m$ transactions, and a new block $b$ carries $n$ \hashes. $M$ may in advance contain any of the transactions that $b$ intends to carry, which will influence the value of $P_{sc}$. But in extreme case $M$ contains none of the transactions, which will lead to the upper bond of $P_{sc}$. In this case the probability of a single collision is approximated as:

\begin{equation}\label{equ:collision}
  P_{sc}=1 - \left[1 - {\left(\frac{1}{2}\right)}^{k}\right]^{\left( mn + \frac{n^{2}}{2} \right)}
\end{equation}

\subsection{Protocol Optimization}
\label{SubSec:sort}

\name can be further optimized to reduce the probability of hash collisions. In this optimization, both the \hashes in a new block and the transactions in the memory pools are sorted by the lexicographic order of \ids. Given a \hash $h$ in the block $b$ with \id = $id$, the candidate space of $h$ will be narrowed to a range bound by $id$ and its previous adjacent \id in $b$, instead of the entire memory pool. This dramatically reduces the collision probability and the cost of resolving ambiguity, which allows even shorter hash with higher compression ratio.

We also use $P_{sc}$ to evaluate the effect of this optimization. If each memory pool contains $m$ transactions, and $n$ \ids in a new block are evenly distributed in the sorted memory pool, the size of the collision space will be reduced from $m$ to $\frac{m}{n}$. The model for $P_{sc}$ is therefore updated as:

\begin{equation}\label{equ:sort}
  P_{sc}=1 - \left[1 - {\left(\frac{1}{2}\right)}^{k}\right]^{ m }
\end{equation}

\subsection{Simulation}
\label{SubSec:Simulation}

We simulated the hash collisions to evaluate above probability models. Sets $M$ and $b$ were defined to simulate the memory pool and new block respectively. We assume $\left|M\right|= m$ and $\left|b\right|= n$. Our simulation set $m = 1000$, and $k$ was iterated from 20 to 40. During each iteration of $k$, we simulated the computation and packing process of \hashes for $N = 100,000$ times, and counted $T_k$, the time of at least one single collision occurring between $M$ and $b$ or within $b$. $P_{sc}$ is approximated as $\frac{T_k}{N}$.

\begin{figure}[htb]
\begin{center}
\begin{tikzpicture}[scale=0.8]
\begin{axis}[
    xlabel={$k$},
    ylabel={$P_{sc}$},
    xmin=20, xmax=35,
    ymin=0, ymax=0.1,
    ymode=log,
    legend pos=north east,
    ymajorgrids=true,
    grid style=dashed,
]

\addplot[
    line width=0.5mm,
    color=red,
    ]
    table[x=k,y=simulation_noctor]
    {n=100.dat};
    \addlegendentry{n=100}

\addplot[
    line width=0.5mm,
    color=blue,
    ]
    table[x=k,y=simulation_noctor]
    {n=300.dat};
    \addlegendentry{n=300}

\addplot[
    line width=0.5mm,
    color=brown,
    ]
    table[x=k,y=simulation_noctor]
    {n=500.dat};
    \addlegendentry{n=500}

\end{axis}
\end{tikzpicture}
\end{center}
 \caption{$P_{sc}$ before Optimization}\label{fig:nCTOR}
\end{figure}
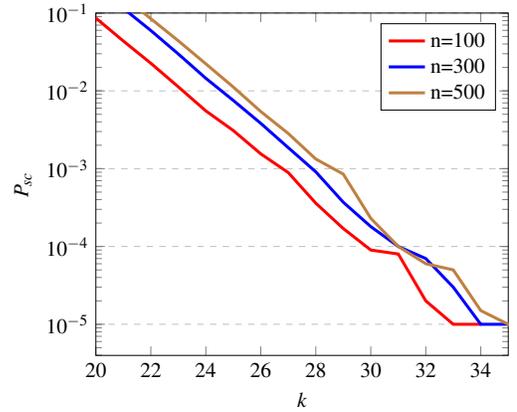

We first simulated the model of $P_{sc}$ under the condition that above optimization was not applied. Figure~\ref{fig:nCTOR} reveals the relation between $P_{sc}$ and $k$ with $n$ = 100, 300, 500 respectively. The maximum value of $k$ on the horizontal axis is 35 because $P_{sc}$ converges to 0 when $k\textgreater35$. For each value of $n$, $P_{sc}$ decreases with the increase of $k$. Furthermore, it demonstrates that $P_{sc}$ increases with the block size $n$, since a large block is more likely to contain \hashes that collide with other ones.

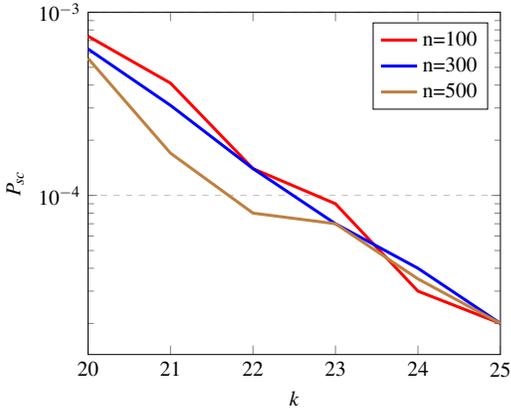
\begin{figure}[htb]
\begin{center}
\begin{tikzpicture}[scale=0.8]
\begin{axis}[
    xlabel={$k$},
    ylabel={$P_{sc}$},
    xmin=20, xmax=25,
    ymin=0, ymax=0.001,
    ymode=log,
    legend pos=north east,
    ymajorgrids=true,
    grid style=dashed,
]

\addplot[
    line width=0.5mm,
    color=red,
    ]
    table[x=k,y=simulation_ctor]
    {ctor_n=100.dat};
    \addlegendentry{n=100}

\addplot[
    line width=0.5mm,
    color=blue,
    ]
    table[x=k,y=simulation_ctor]
    {ctor_n=300.dat};
    \addlegendentry{n=300}

\addplot[
    line width=0.5mm,
    color=brown,
    ]
    table[x=k,y=simulation_ctor]
    {ctor_n=500.dat};
    \addlegendentry{n=500}

\end{axis}
\end{tikzpicture}
\end{center}
 \caption{$P_{sc}$ after Optimization}\label{fig:CTOR}
\end{figure}

Figure~\ref{fig:CTOR} reveals the simulation results after the optimization. This optimization achieves much smaller $P_{sc}$ compared with the results in Figure~\ref{fig:nCTOR}, and $P_{sc}$ converges to 0 when $k\textgreater25$. We recommend $k=32$ as a reasonable value with a better compression ratio and low collision rate. When $m=3000$ and $n=200$, a $32$-bit \hash makes $P_{sc} =0.00000072$. This yields an 80 times of data size reduction compared with the original version of the blockchain.

\section{Collision Attack}
\label{Sec:Attack}

An attacker may create a new transaction with its \hash matched to an existing transaction. A flood of such malicious transactions for collisions can invalidate our collision probability analysis and make the verification of new blocks expensive, which eventually results in higher fork rate.  We propose simple strategies to address this potential attack.

A simple strategy for defense is to introduce a salt while calculating the \hash:

\begin{center}
  \texttt{TXID-HASH} = \texttt{hash( Salt + TXID )}
\end{center}

The salt is specific to the block carrying these \hashes, and is included in the encoded data (e.g., the block header). For example, just take the CRC32-Merkle root as the salt, or introduce another 4-byte field with random bits.

By introducing salts, the attacker cannot create malicious transactions that match existing transactions, until a new block carrying these existing transactions is generated. Malicious transactions are also unlikely to reach the entire network faster than a new block unless the attacker controls a large botnet.

In extreme cases, the massive collision attack is still possible regardless of the high cost. We require miners fall back to \id list when the entire network is under attack. Such attack can be noticed by all miners when verifying new blocks. A miner can simply count the number of ambiguous \hash per-block. If the counts are significantly higher than the expected value and forks are observed, the next block should fall back to \id list.

\section{Conclusion}
\label{Sec:Conclusion}

This paper proposes Txilm, a protocol that compresses transaction representation in the blockchain. Each block in Txilm carries the short hashes of \ids (i.e., \hashes) instead of complete transactions. When receiving a block from the peers, a full node can search the transactions in its local memory pool based on the \hashes and resolve the hash collisions using Merkle root. We analyze the probability of hash collisions by simulation, and use the salted hash to protect against potential attacks. Combined with the sorted transactions based on \ids, Txilm realizes 80 times of data size reduction and dramatically saves the network bandwidth. This will lead to a higher blockchain throughput.

In our future work, we will implement Txilm Protocol, and evaluate the effect of this protocol on the blockchain throughput by experiments. 
\bibliographystyle{plain}
\bibliography{main}
\end{document}